\documentclass[12pt,letterpaper]{article}
\usepackage{setspace}
\usepackage{color}
\usepackage{amsmath}
\usepackage{amsfonts}
\usepackage{verbatim}
\usepackage{amssymb}
\usepackage{graphicx,bm}
\usepackage{graphicx}
\usepackage{bbm}
\usepackage{cite}
\usepackage{authblk}
\usepackage[shortlabels]{enumitem}

\newcommand{\ima}{\mathbbmtt{i}}

\numberwithin{equation}{section}

\author[1]{N. Dimakis\thanks{nsdimakis@scu.edu.cn, nsdimakis@gmail.com}}
\affil[1]{Center for Theoretical Physics, College of Physical Science and Technology Sichuan University, Chengdu 610065, China}
\author[2]{Petros A. Terzis\thanks{pterzis@phys.uoa.gr}}
\author[2]{T. Christodoulakis\thanks{tchris@phys.uoa.gr}}
\affil[2]{Nuclear and Particle Physics Section, Physics Department, National and Kapodistrian University of Athens, GR 157–71 Athens, Greece}

\begin{document}

\title{\textbf{Integrability of geodesic motions in curved manifolds through
non-local conserved charges}}
\date{}
\maketitle
\begin{abstract}
In this work we study the general system of geodesic equations for the case of a massive particle moving on an arbitrary curved manifold. The investigation is carried out from the symmetry perspective. By exploiting the parametrization invariance property of the system we define nonlocal conserved charges that are independent from the typical integrals of motion constructed out of possible Killing vectors/tensors of the background metric. We show that with their help every two dimensional surface can - at least in principle - be characterized as integrable. Due to the nonlocal nature of these quantities not more than two can be used at the same time unless the solution of the system is known. We demonstrate that even so, the two dimensional geodesic problem can always be reduced to a single first order ordinary differential equation; we also  provide several examples of this process.
\end{abstract}

\section{Introduction}

The existence of symmetries is of paramount importance in all aspects of physical theories \cite{Sundnew}. The search of analytic solutions in mechanical systems is greatly facilitated when first order relations provided by existing integrals of motion are present. The notion of integrability itself is related to the existence of enough independent, commuting phase space functions that are constants of motion \cite{Arnold,Cam}. In this work we use this term to refer to Liouville integrability and the Liouville-Arnold theorem \cite{Arnold}.

The present  work is devoted to the study of a particular class of mechanical systems. Namely, those that describe the motion of relativistic free particles in (generally) curved spaces, i.e. geodesic problems (the inclusion of a potential however, does not essentially affect the analysis  we follow; we are going to see this in a later section). The study of geodesic equations and their symmetries is a subject extensively examined in the literature \cite{Eisenhart,Carter,Hojman,Bolsinov,Benn,KYan,Andr1,Eva,new1,new2}. Especially in the case of pseudo-Riemannian spaces where it is of particular importance in gravitational problems \cite{Camci,Bocca,Frolov,Page}. Apart from the case where the metric describes the base manifold of some space-time in a gravitational theory, a pseudo-Riemannian geodesic problem may also be related with systems appearing in the context of mini-superspace cosmology \cite{tchris1,tchris2}. The latter are also parametrization invariant and are completely equivalent to some geodesic problem as we shall see later on in the analysis. A similar correspondence can be claimed for regular (non-parametrization invariant) mechanical systems up to a transformation in time with the use of the Jacobi metric \cite{Arnold,Benn,Awre,Gibbons}. An alternative way of a geometrization of classical regular mechanical problems is supplemented by the Eisenhart-Duval lift \cite{Eisenhart2,Duv1,Duv2,Eis3} which was initially introduced by Eisenhart and later rediscovered in a more physical context by Duval and collaborators; for a recent treatment of two dimensional problems see \cite{Fordy}. Thus, we see that the motion of a relativistic particle in a curved manifold can be associated with several different problems. In what regards in particular the motion of a such a particle in a Minkowski space and under several different contexts we refer to the study presented in \cite{Chile}.

In this work we choose to follow a different procedure than what is frequently encountered in the literature. The starting point of a geodesic problem is usually the set of equations
\begin{equation}\label{affgeo}
  \ddot{x}^\mu + \Gamma^\mu_{\kappa\lambda} \dot{x}^\kappa \dot{x}^\lambda = 0,
\end{equation}
where the $x^\alpha$'s are the local coordinates on a given manifold and $\Gamma^\mu_{\kappa\lambda}$ the Christoffel symbols corresponding to its metric $g_{\beta\gamma}(x)$. The set of equations \eqref{affgeo} describes however a very particular type of geodesics; those that are characterized by an affine parameter. Here, we choose to study the more general system
\begin{equation}\label{gengeo}
  \ddot{x}^\mu + \Gamma^\mu_{\kappa\lambda} \dot{x}^\kappa \dot{x}^\lambda = \frac{1}{2}\dot{x}^\mu \frac{d}{d\tau} \left[ \ln \left(\dot{x}^\kappa \dot{x}_\kappa \right) \right],
\end{equation}
which is the corresponding set of the full  geodesic equations (prior to fixing the gauge, e.g. by deciding that the parameter along the integral curves is affine). As we shall see, the parametrization invariance which is inherent in the action principle producing \eqref{gengeo}, allows for the definition of additional symmetries leading to rheonomic conserved charges. These are supplementary to the conventional integrals of motion that may appear when one concentrates on the gauged fixed Lagrangian of the geodesic problem. Although these new conserved charges  have a nonlocal form, they can, under specific gauge choices, provide additional first order relations that may help to integrate particular systems under consideration. These integrals of motion are not of ``Noetherian" origin, i.e. they are not the result of an existing variational symmetry, not even of a generalized one. However, their realization is inextricably linked to the parametrization invariance.

In our analysis we provide a method for using such, supplementary, first order relations in order to integrate \eqref{gengeo}. One of the main points we wish to make is that equations \eqref{gengeo} may be a lot easier to solve under a smart gauge fixing choice, than the ``seemingly" simplest \eqref{affgeo}. Furthermore, we study in what way the integrability of such a system may be affected by the use of the nonlocal conserved charges and we use one such integral of motion in order to reduce the problem of the geodesic motion on a general two dimensional manifold to a single first order differential equation. Something which we show that it is in principle always possible, assuming that the metric is smooth enough.

The structure of the paper is as follows: In section 2 we provide the general setting of the geodesic problem and review some basic properties of the associated systems of equations. In section 3 we turn to the phase space description where we introduce the nonlocal conserved charges that can be used together with the conventional integrals of motion. In section 4 we study the ways in which the former may affect the notion of integrability in a given system. Section 5 is devoted to the reduction of the general two-dimensional geodesic problem with the help of such a nonlocal conserved charge. In section 6, we briefly review some examples and applications of the general form of relations derived in the previous section. Finally, our conclusions are given in section 7.

\section{The geodesic problem}

Let us consider the problem of deriving the trajectory of a free particle of unit mass  moving in a space of dimension $d$ characterized by a metric with components $g_{\mu\nu}(x)$. The usual starting point in the literature is the well known Lagrangian
\begin{equation}\label{Lagreg}
  L_1=\frac{1}{2} g_{\mu\nu} \dot{x}^\mu \dot{x}^\nu , \quad\quad \mu,\nu =1,...,d,
\end{equation}
with $x^\mu$ being the coordinates on the manifold and $\dot{x}^\mu$ their first derivatives with respect to some parameter $\tau$ in terms of which the trajectory is to be described, i.e. $\dot{x}^\mu= \frac{d}{d\tau}x^\mu(\tau)$. However, Lagrangian \eqref{Lagreg} does not tell us the full story. It reproduces as solutions to its equations of motion only a very specific type of geodesic trajectories;  those associated with an affine parameter.

The full geodesic equations \eqref{gengeo}, in an arbitrary parametrization, are given by the square root Lagrangian
\begin{equation}\label{Lagsq}
  L_2 = \sqrt{|g_{\mu\nu} \dot{x}^\mu \dot{x}^\nu |} .
\end{equation}
Strictly speaking, the two systems described by $L_1$ and $L_2$ are not completely equivalent. The basic difference being that the latter is parametrization invariant. The ``time" parameter $\tau$ in the system described by $L_1$ has the status of a Newtonian-like time, in the sense that the system exhibits a Noether symmetry generated by $X_1 = \frac{\partial}{\partial \tau}$; while the action constructed with the help of $L_2$ possesses a symmetry generator of the form $X_2 = F(\tau)\frac{\partial}{\partial \tau}$ with $F(\tau)$ being an arbitrary function.\footnote{The generator $X_1$ implies that a system is invariant under constant translations in time $\tau\rightarrow \tau+ c$, while $X_2$ allows for time parametrization invariance $\tau \rightarrow f(\tau)$. In the first case, constant translations in time are the ``gauge" transformations for the system. If one had performed any other time transformation it would result in affecting the physical properties of the latter.} The infinite dimensional symmetry group available in the second case implies - in place of a conservation law - the existence of a differential identity among the equations of motion (second Noether theorem). In other words that not all of them are independent.

Lagangian $L_2$ is homogeneous of degree one in the velocities, i.e.
\begin{equation}\label{homogeneous}
  L_2 (x,\lambda \dot{x}) = |\lambda|  L_2 (x,\dot{x})
\end{equation}
and this results in the Hamiltonian $H$ being identically zero. This is a consequence of Euler's theorem for homogeneous functions, which implies in this case that \cite{Gourg}
\begin{equation}
  \frac{\partial L_2}{\partial \dot{x}^\kappa} \dot{x}^\kappa - L_2 \equiv 0.
\end{equation}
In the left hand side we recognize the Hamiltonian, $\frac{\partial L_2}{\partial \dot{x}^\kappa}\equiv p_\kappa$ being the momenta conjugate to $x^\kappa$. Thus, one obtains $H \equiv 0$. However, this complication can be avoided by using another Lagrangian dynamically equivalent to $L_2$ which has an additional, auxiliary degree of freedom that we denote with $N$. This Lagrangian is written as
\begin{equation}\label{Lag}
  L_3 = \frac{1}{2N} g_{\mu\nu} \dot{x}^\mu \dot{x}^\nu - \frac{N}{2}, \quad\quad \mu,\nu =1,...,d,
\end{equation}
and the $N(\tau)$ is to be considered as a degree of freedom on an equal footing with the $x(\tau)^\alpha$'s. In other words the system described by $L_3$ - in contrast to those of $L_1$ and $L_2$ - has $d+1$ degrees of freedom instead of just $d$. We can thus write $L_3 = L_3 (q,\dot{q})$, where $q^i=(q^\mu,q^{d+1})=(x^\mu,N)$, $i=1,...,d+1$. The equations of motion for $L_3$ can be easily derived and they result in the following system of ordinary differential equations
\begin{subequations}\label{Eul}
  \begin{align} \label{EulN}
    \frac{\partial L_3}{\partial q^{d+1}}- \frac{d}{d\tau} \left( \frac{\partial L_3}{\partial \dot{q}^{d+1}} \right) & = 0 \Rightarrow \frac{1}{N^2} g_{\mu\nu} \dot{x}^\mu \dot{x}^\nu+1  =0 \\ \label{Eulx}
    \frac{\partial L_3}{\partial q^{\mu}}- \frac{d}{d\tau} \left( \frac{\partial L_3}{\partial \dot{q}^{\mu}} \right) & = 0 \Rightarrow \ddot{x}^\mu + \Gamma^\mu_{\kappa\lambda} \dot{x}^\kappa \dot{x}^\lambda -\dot{x}^\mu \frac{d}{d\tau}\left( \ln N\right)  =0,
  \end{align}
\end{subequations}
where of course if you solve \eqref{EulN} algebraically with respect to $N$ and substitute into \eqref{Eulx} you obtain  the set of equations \eqref{gengeo} (in this situation we have written $L_3$ in such a manner so as to consider a metric with Lorentzian signature in which, for time-like geodesics, $\dot{x}^\kappa \dot{x}_\kappa <0$ holds\footnote{If for example we wanted to treat a Riemannian metric (or space-like ``trajectories") we should write $L_3$ with $+\frac{N}{2}$ in the potential part}). Hence, the two Lagrangians, $L_2$ and $L_3$, are equivalent and possess exactly the same symmetry group as both actions constructed by them are parametrization invariant, i.e. both of them admit $X_2$ as a Noether symmetry generator. The great difference however is that the Hamiltonian corresponding to $L_3$ is not identically zero as the one of $L_2$ but rather weakly zero, thus allowing phase space dynamics. The additional degree of freedom $N$, is referred in the literature as the einbein \cite{Brink,Sundnew} and is usually symbolized with an $e$. Here, we prefer to use $N$ due to the relation it bears with the lapse function of mini-superspace cosmological models and the Lagrangians that emerge in that context.

Due to the parametrization invariance of the system characterizing $L_3$ we have the freedom to fix the gauge in the set of equations \eqref{Eul}. Obviously the choice $N=$const. leads us to the affinely parameterized geodesic equations \eqref{affgeo}. This is a seemingly opportune choice in order to simplify the system at hand. However, we are going to demonstrate here that the set of equations \eqref{Eul} may be a lot easier to solve if a gauge choice smarter than the obvious $N=$const is employed.

Before we proceed let us demonstrate how one may be led to \eqref{Lag} from a more general problem of a particle moving under the influence of some potential $V(x)$. In this case the Lagrangian would read
\begin{equation}\label{LagV}
  L_4 = \frac{1}{2n} \bar{g}_{\mu\nu} \dot{x}^\mu \dot{x}^\nu - n V(x) .
\end{equation}
These type of Lagrangians appear in the study of cosmological systems as mini-superspace models, whenever they happen to reproduce correctly the field equations under some ansatz for the base manifold metric. The $n$ in this case is usually the lapse function of the latter. By performing a simple rescaling of the degree of freedom $n\mapsto N = 2 n V$ we obtain $L_3$ where $g_{\mu\nu}= 2 V(x) \bar{g}_{\mu\nu}$, i.e. a geodesic problem of a conformally related metric with the potential serving as the conformal factor. A similar identification can be done for regular systems without constraints and the resulting metric $g_{\mu\nu}$ is called the Jacobi metric \cite{Arnold,Benn,Awre}. The only difference there is that the equivalence applies modulo a time transformation. This is due to the fact that the aforementioned scaling of $N$ is equivalent to making a change in time $\tau \mapsto t = 2 \int V(x(\tau)) d\tau$ in the action. For a parametrization invariant system this is just a change in gauge, for a regular one it is a transformation that does not belong to its symmetry group. Hence, it alters the physical properties. Nevertheless, this is not of essence when the objective is the integrability; the obtained solution can always be mapped to the original system through the inverse transformation. There is an immense bibliography on symmetries of mini-superspace systems, for the interested reader we just refer to a few characteristic  \cite{Cap,Vakili,Cap2,Terzis,Andr2,Andr4,Dim1,Sharif,Cap3,Andr3}.

\section{Phase space description and a new class of conserved quantities}

\subsection{Phase space of a singular Lagrangian}

A Lagrangian function $L(q,\dot{q})$, where with $q$ we designate the generalized coordinates of the configuration space, is considered to be constrained (or singular) if the corresponding Hessian matrix $W_{ij}=\frac{\partial^2 L}{\partial \dot{q}^i \partial \dot{q}^j}$  has a zero determinant. This results in the Legendre transform not being invertible, which poses an issue for passing to the Hamiltonian description. Dirac \cite{Dirac1,Dirac2,Dirac3} and Bergmann \cite{Berg} separately provided a solution to this problem. Before proceeding with the application of this formalism on our particular case of the geodesic Lagrangian, let us make first a brief introduction to the Dirac-Bergmann algorithm (for more details we refer the interested reader to several textbooks that exist on the subject \cite{Dirac3,conbook1,conbook2,conbook3}). In what follows we mostly follow the presentation of the theory carried out in \cite{conbook1}.

Assume a Lagrangian function  possessing $D$ degrees of freedom $q^i$, $i=1,...,D$, (in the case of $L_3$ in \eqref{Lag} $D=d+1$) and which satisfies the singularity condition
\begin{equation}
 \mathrm{Det}(W_{ij})=\mathrm{Det}\left(\frac{\partial^2 L}{\partial \dot{q}^i \partial \dot{q}^j} \right) = 0, \quad i,j=1,...D.
\end{equation}
This implies that $\mathrm{Rank}(W_{ij})=R<D$ and leads to the existence of a set of $D-R$ equations in number which do not contain accelerations. Thus, not all of the equations of motion are independent.

For clarity, let us split the index $i=1,...,D$, which indicates all the degrees of freedom, into two sub-indexes $\alpha$ and $I$. The first, $\alpha=1,...,R$, corresponds to the degrees of freedom of the invertible sub-matrix of $W_{ij}$ and the second, $I=R+1,...,D$, to those that are left. When a variable appears with no index in an argument of a function, we assume the full range i.e. $f(x)$ means $f(x^i)$, $i=1,...,D$.

We can normally define the momenta of the system as $p_i = \frac{\partial L}{\partial \dot{q}^i}$ for all the range of $i=1,...,D$. For $R$ of the velocities of the system, say $\dot{q}^\alpha$, there exists a one to one correspondence with the equal in number momenta $p_\alpha$, $\alpha=1,...,R$. The rest of the momenta, $p_I$ with $I=R+1,...,D$, are given in terms of $x$ and $p_\alpha$. These $D-R$ in number relations among position and momenta, that involve no velocities, are called the primary constraints of the system, which we can denote as
\begin{equation} \label{primary}
  \phi_I (q,p) = 0 , \quad I = R+1,...,D.
\end{equation}
Even though the Legendre transform is no longer invertible, one can write a Hamiltonian function $H_c = p_i \dot{q}^i -L$, where the $p_I$, which are not associated with any of the velocities, it is understood that they are being substituted from \eqref{primary}. This Hamiltonian is defined in the space of the $D$ in number $q^i$, the $R$ momenta $p_\alpha$ - that do have an one to one correspondence with velocities - and the rest $D-R$ velocities $\dot{q}^I$. The physical space of the problem is the subspace where the relations \eqref{primary} hold. As it turns out, the Hamiltonian $H_c = p_i \dot{q}^i -L$, is not uniquely defined since one can add any linear combination of the constraints \eqref{primary} which are zero \cite{Dirac3}. Thus we can write the general Hamiltonian
\begin{equation} \label{primaryham}
  H = H_c + u^I \phi_I,
\end{equation}
where the $u^I$ are functions that may depend on $q$ and $p$. The part of the primary Hamiltonian that is denoted with $H_c$ is called the canonical Hamiltonian. Poison brackets can be defined normally in the space of $q^i$ and $p_i$, i.e. $\{A,B\} = \frac{\partial A}{\partial q^i}\frac{\partial B}{\partial p_i}-\frac{\partial A}{\partial p_i}\frac{\partial B}{\partial q^i}$ and with their help the evolution of phase space quantities with respect to $H$ can be calculated. However, it is important to notice that the quantities $\phi_I=0$ are not to be set to zero prior to carrying the full Poisson bracket calculation; else the result is erroneous. This rule defines the notion of the weak equality which is denoted by an ``$\approx$", hence from now on we write the constraint equations as $\phi_I \approx 0$.\footnote{Formally, a weak equality for a quantity $\phi_I\approx 0$ means that it is zero itself but its gradient (in phase space) is not. This is why it is important not to put $\phi_I$ equal to zero inside Poisson brackets, since the latter produce terms involving $\frac{\partial \phi_I}{\partial q^i}$ and $\frac{\partial \phi_I}{\partial p_i}$ which are the components of the aforementioned gradient.}

Of course, due to the fact that through the dynamical evolution one is not allowed to leave the physical space, the time evolution of the constraints must vanish at least on the constrained surface itself. In other words we have to impose the supplementary condition
\begin{equation} \label{consistency}
  \dot{\phi}_I = \{\phi_I,H\} \approx 0.
\end{equation}
Each of these additional equations may bring about the following results:
\begin{enumerate}[a)]
  \item It can be satisfied identically.
  \item Lead to a new relation among position and momenta that have to vanish weakly. That is, to a new constraint $\psi(q,p)\approx 0$ which is called secondary. \label{consistencyb}
  \item Define one of the functions $u_I$. In this case the \eqref{consistency} which results to
  \begin{equation} \label{uJsol}
    \{\phi_I, H_c\} + u^J \{\phi_I,\phi_J\} \approx 0 .
  \end{equation}
  can be solved with respect to some of the multipliers $u^I$.
  \item Lead to an inconsistency. This case emerges if the action of the system has no extremum i.e. the Euler-Lagrange equations are incompatible.
\end{enumerate}
In the case \ref{consistencyb}, where secondary constraints emerge, the initial physical space defined by $\phi_I \approx 0$ has to be further restricted by any additional secondary constraint $\psi_\mathcal{I}\approx 0$, with $\mathcal{I}$ the index counting the number of second class constraints. Of course the consistency condition $\dot{\psi}_\mathcal{I}\approx 0$ has also to be imposed for each of the secondary constraints, which may lead in its turn to any of the four previously described possibilities. If the system is consistent, after a finite number of steps, we arrive to a situation where the process closes without the emergence of any new secondary constraints\footnote{Here ee use the term secondary to describe all constraints that are not primary, irrespectively of the number of times the eventuality \ref{consistencyb} may emerge.}.

After having calculated all the constraints of our theory, a very important distinction takes place. We split them into first and second class. The first class constraints are defined as those that commute at least weakly through the Poisson bracket with all the rest of the constraints, while on the other hand the second class are those that do not\footnote{In the process of making this distinction, a linear rearrangement of the constraints might be necessary so that we obtain the maximum number of first class constraints that are present in the given system}. We decide to make the following distinction and denote first class constraints with the sub-indexes ``$f$" and ``$s$" respectively. We have thus the primary constraints $\phi_I$ split into the subsets $\phi_{I_{f}}$ and $\phi_{I_{s}}$ depending on whether they are first or second class respectively. The different indexes $I_f$ and $I_s$ indicate that each one runs in the range of the cardinality of each subclass. By making the same distinction for the second class constraints we can divide $\psi_\mathcal{I}$  into $\psi_{\mathcal{I}_{f}}$ and $\psi_{\mathcal{I}_{s}}$ again depending on if they are first or second class.

In the end we can write the Hamiltonian of the system from \eqref{primaryham} as
\begin{equation}
  H = H_c +u^{I_{f}} \phi_{I_{f}} + u^{I_{s}} \phi_{I_{s}},
\end{equation}
where $u^{I_{f}}$ is the subset of the functions $u^I$ in \eqref{primaryham} that remain arbitrary due to the fact that $\phi_{I_{f}}$ as first class commute with all the rest of the constraints, while $u^{I_{s}}$ are those of the $u^I$ which have been obtained in terms of $q$ and $p$ from \eqref{uJsol}. The latter coefficients can be written explicitly, if we put all the second class constraint in one set denoted by $\zeta_\Gamma = (\phi_{I_{s}}, \psi_{\mathcal{I}_{s}})$, where the capital Greek letter $\Gamma$ runs as an index through the combined range of $I_s$ and $\mathcal{I}_s$. We can now define the invertible matrix (note that the number of second class constraints is always even \cite{Dirac3})
\begin{equation} \label{deltamat}
  \mathcal{C}_{\Gamma\Delta} = \{\zeta_\Gamma,\zeta_{\Delta}\}.
\end{equation}
Then, equations \eqref{uJsol} lead to
\begin{subequations} \label{solveu}
\begin{align}
  u^{I_{s}} = - \left(\mathcal{C}^{-1}\right)^{I_{s} \Gamma} \{\zeta_\Gamma, H_c\} \\
  \left(\mathcal{C}^{-1}\right)^{\Gamma\Delta }\{\zeta_\Delta , H_c\} \approx 0,
\end{align}
\end{subequations}
where $\mathcal{C}^{-1}$ is the inverse of the matrix \eqref{deltamat}, i.e. $\left(\mathcal{C}^{-1}\right)^{\Gamma\Delta }\mathcal{C}_{\Delta\Theta}= \delta^{\Gamma}_{\Theta}$, with $\delta$ denoting the usual Kronecker delta. Finally, with the help of \eqref{solveu} - and by using the fact that any second degree expression in the constraints is strongly zero - the dynamical evolution of any quantity $A(t,q,p)$ is given by
\begin{equation}
  \dot{A} = \frac{\partial A}{\partial t} + \{A,H\} = \frac{\partial A}{\partial t} + \{A,H_c\} + u^{I_{f}} \{A, \phi_{I_{f}}\} - \{A,\zeta_\Gamma \} \left(\mathcal{C}^{-1}\right)^{\Gamma \Delta}\{\zeta_{\Delta}, H_c\}.
\end{equation}
The first class constraints are usually related to some existing gauge freedom in the problem under consideration, while the second class constraints simply denote redundant degrees of freedom.

In the case of the geodesic system, as we are going to see next, the procedure leads to a situation described purely by first class constraints. However, the introduction of a gauge fixing condition as a supplementary constraint turns the existing primary first class constraint into a second class. In systems where second class constraints are present the dynamical evolution can be given with the help of the Dirac bracket which is defined as
\begin{equation}\label{Dbradef}
  \{\;,\; \}_D = \{\;,\; \}  - \{\; ,\zeta_\Gamma \} \left(\Delta^{-1}\right)^{\Gamma \Delta}\{\zeta_{\Delta},\;  \}.
\end{equation}
If one has only second class primary constraints, the evolution in time of a quantity $A(t,q,p)$ reads
\begin{equation}
  \dot{A} = \frac{\partial A}{\partial t} + \{A,H_c\}_D .
\end{equation}
What is more, due to the fact that $\{A, \zeta_\Gamma\}_D \equiv 0$, the distinction between a weak and a strong equality for the second class constraints is no longer necessary. Thus, allowing the elimination of the overabundant degrees of freedom. Apart from the aforementioned textbooks, for further discussions on the Dirac brackets we refer to \cite{DB1,DB2}.

Those are all the theoretical tools we need at our disposal. We may now proceed and put them in practice for the geneal geodesic Lagrangian under consideration.

\subsection{Hamiltonian description for the geodesic problem}

We can straightforwardly observe that the Lagrangians which we want to study - namely $L_2$ and $L_3$ - are constrained. The Hessian matrices $\frac{\partial^2 L_2}{\partial {\dot{x}}^\mu \partial \dot{x}^\nu}$ of the first and $W_{ij}=\frac{\partial^2 L_3}{\partial {\dot{q}}^i \partial \dot{q}^j}$ of the second (where $q^i=(x^\mu,N)$, $i,j=1,...,d+1$) have zero determinants. We already discussed that the $L_2$ system has an identically zero Hamiltonian. Hence, we concentrate our attention on the Lagrangian $L_3$.

By following the Dirac-Bergmann prescription, which we described in the previous subsection, we first have to identify the constraints of the system; that is the relations among momenta and positions that do not involve velocities. For $L_3$ we can immediately see that such a constraint is the relation
\begin{equation}
  p_N = \frac{\partial L_3}{\partial \dot{N}} = 0,
\end{equation}
which is zero due to the fact that, in $L_3$, there is no velocity for the degree of freedom $q^{d+1}=N$. The above is the primary constraint of the system and is denoted as $p_N \approx 0$.\footnote{Here we can immediately see why the notion of a weak equality is important. Obviously, the derivative of $p_N$ with respect to itself is not zero and thus its gradient in phase space is not vanishing. So, it would be erroneous to set $p_N$ equal to zero before Poisson brackets are calculated. It is correct to write $\{N,p_N\}=1$, but it wrong to set $\{N,p_N\}=\{N,0\}=0$.} We can also note that this is the only primary constraint since the rank of the $W_{ij}$ matrix is $R=d=\mathrm{dim}(g_{\mu\nu})$ - of course we assume that the metric $g_{\mu\nu}$ in $\eqref{Lag}$ is invertible. According to the discussion of the previous section we expect $D-R$ in number primary constraints, which in this case is $D-R=d+1-d=1$. We can now proceed by writing down the Hamiltonian which is
\begin{equation} \label{Ham}
  H = p_i \mathrm{\dot{v}}^i - L_3 = p_N \dot{N} +p_\mu \dot{q}^\mu -L_3 = \frac{N}{2} \left(g^{\mu\nu} p_\mu p_\nu+1\right) + p_N u_N,
\end{equation}
where we decided to write $\dot{N} = u_N$. In comparison to relation \eqref{primaryham} of the previous section we observe that the canonical Hamiltonian corresponds to the first part of \eqref{Ham}, while the primary constraint $\phi_1=p_N$ appears with a multiplier in the Hamiltonian.

We now need to satisfy the consistency condition \eqref{consistency}, that the constraint is preserved (at least weakly) through the time evolution. As we discussed the system cannot leave the constrained surface, which is identified as the physical space. Hence, we need to demand
\begin{equation}
  \dot{p}_N \approx 0 \Rightarrow \{p_N, H\} \approx 0 \Rightarrow  -\left(g^{\mu\nu} p_\mu p_\nu+1\right) \approx 0 .
\end{equation}
The quantity
\begin{equation}
  \mathcal{H} := g^{\mu\nu} p_\mu p_\nu+1 \approx 0
\end{equation}
is the secondary constraint (in the general formulation of the previous section $\psi_1=\mathcal{H}$), whose conservancy through time does not lead to any tertiary constraints since
\begin{equation}
  \dot{\mathcal{H}} = \{\mathcal{H}, H\} = 0 .
\end{equation}
The process terminates here and we obtain Hamiltonian \eqref{Ham} as a linear combination of constraints. The latter are both categorized as first class constraints\footnote{Remember that a first class constraint is one that commutes (at least weakly) with all the rest of the constraints.} since $\{p_N, \mathcal{H}\} =0$. Due to this, the multiplier $u_N$ remains an arbitrary function whose value is not determined by the evolution of the system.

The notion of weak equality in constrained systems allows us to extend what we may consider as a conserved charge in phase space. Kucha\v{r} was the first to introduce what we call a conditional symmetry \cite{Kuchar}. This is defined as a quantity, linear in the momenta, which is conserved due to the Hamiltonian constraint (for more recent applications of quantities which are conserved on the constrained surface see \cite{tchris1,Igata0}). Assume that $Q(x,p)$ is such a quantity; then if
\begin{equation}\label{prop1}
  \{Q,\mathcal{H}\} = \tilde{\omega}(x) \mathcal{H},
\end{equation}
where $\tilde{\omega}$ is some function of the configuration space variables, then this $Q$ has the property of been conserved on the constrained surface, i.e. $\frac{d Q}{ d \tau} \approx 0$. Obviously $\frac{d Q}{ d \tau} \approx 0$ is less restrictive than $\frac{d Q}{ d \tau} = 0$\footnote{If we see \eqref{prop1} as a partial differential equation for $Q(x,p)$, then $\{Q,\mathcal{H}\}=0$ is the corresponding homogeneous equation. The solution of the latter is of course contained in the general solution of \eqref{prop1}.}, thus in constrained systems you may have a larger symmetry group than what is encountered in classical regular systems.

In the case of a geodesic problem, which is described by \eqref{Lag} and in phase space by \eqref{Ham}, the consideration of a quantity $Q=\xi^\alpha (x) p_\alpha$ implies the property \eqref{prop1} if the $\xi^\alpha$'s form the components of a Killing vector field of metric $g_{\mu\nu}$ (for $\tilde{\omega}=0$ in \eqref{prop1}). The same is also true for higher order symmetries constructed out of Killing tensors of $g_{\mu\nu}$, as it is well known in the literature. So, we see that - in terms of a geodesic problem - there is no difference between the regular system \eqref{Lagreg} and the geodesic systems $\eqref{Lagsq}$ and \eqref{Lag} in what regards these particular types of symmetries. Of course, in the case where one considers null geodesics this group expands and conformal Killing tensors or vectors can be used to construct local conserved charges. The latter property of null geodesics was used recently in the realization of a conserved charge out of the scaling symmetry of the Kepler system \cite{Horv2}. In what our regards our study, and the motion of a massive particle, we shall immediately see that we can find additional symmetries if we try to extend the original notion of a conditional symmetry as introduced by Kucha\v{r}.

To that end, let us consider a quantity
\begin{equation}\label{nonloc1}
  I = \xi^\alpha(x) p_\alpha + A(\tau)
\end{equation}
which involves a part containing an explicit dependence on time. The demand that its time derivative vanishes weakly leads to
\begin{equation}
  \frac{d I}{d\tau} \approx 0\Rightarrow \frac{d I}{d\tau} =N \omega (x) \mathcal{H} \Rightarrow \frac{d A}{d \tau} + \{\xi^\alpha(x) p_\alpha, H\} = N \omega (x) \mathcal{H}.
\end{equation}
In the right hand side we have chosen to substitute the zero in the weak equality with an expression linear in the quadratic constraint.\footnote{Given the quantities of the left hand side this is the only possibility for the right hand side.} If $\xi$ is a conformal Killing vector of the metric $g_{\mu\nu}$ with conformal factor $2\omega(x)$, i.e. $\mathcal{L}_\xi g_{\mu\nu}=2 \omega g_{\mu\nu}$, where $\mathcal{L}$ denotes the Lie derivative, then we obtain
\begin{equation}
  \frac{d A}{d\tau} = N(\tau) \omega (x(\tau)) .
\end{equation}
Hence, if the quantity $A(\tau)$ is such that
\begin{equation}
  A(\tau) = \int\!\! N \omega d\tau,
\end{equation}
then $I$ of \eqref{nonloc1} will be conserved on the constrained surface. As a result, for every conformal Killing vector $\xi$ of $g_{\mu\nu}$ with conformal factor $2 \omega$ we have a conserved quantity which in general has the nonlocal form \cite{tchris1}
\begin{equation} \label{nonloc}
  I = \xi^\alpha(x) p_\alpha + \int\!\! N \omega d\tau,
\end{equation}
due to involving a part that is a time integral of functions of phase space variables. Of course, when $\xi$ is a Killing vector field ($\omega=0$), we have the usual local conserved charge that is just linear in the momenta.

Such a conserved charge like \eqref{nonloc} defines an additional supplementary equation that holds on mass shell, $I=0$.\footnote{Note that you do not need to consider $I=const.\neq 0$ because expression \eqref{nonloc} involves an indefinite integral which already entails an arbitrary constant of integration.} It can be argued that such an equation is of no great use due to being an integrodifferential equation. However, we have to note that up to this point we have not exploited the freedom of fixing the gauge. If we choose the gauge in such a manner so that $N= \frac{1}{\omega}$, then equation \eqref{nonloc} leads to the first order differential equation (after the momenta are expressed in terms of velocities $p_\alpha \equiv \frac{1}{N} g_{\mu\alpha}\dot{x}^\mu$)
\begin{equation} \label{extra}
  \xi^\alpha(x) p_\alpha + \tau + c = 0,
\end{equation}
where $c$ is a constant of integration. The latter, without any loss of generality, can be set to zero since it can be absorbed with a transformation $\tau\rightarrow \tau - c$. Note that this equation is necessarily functionally independent from any other first order relation constructed by local conserved charges (or the one given by the constraint itself $\mathcal{H}=0$) since $\tau$ appears explicitly in it. Equation \eqref{extra} can be used in order to obtain the solution of the system, because it supplies us with the information that, in the gauge $N= \frac{1}{\omega}$, the combination $\omega(x(\tau)) \xi_\alpha(x(\tau))\dot{x}^\alpha(\tau)$ is equal to $-\tau$.

Another way to exploit an integral of motion like \eqref{nonloc} is to parametrize the einbein field as $N=\frac{\dot{h}(\tau)}{\omega}$. This leads $I$ into becoming
\begin{equation} \label{repar}
  I = \xi^\alpha(x) p_\alpha + h(\tau) + c = 0
\end{equation}
Once more we may set $c=0$, due to the freedom of selecting at will the function $h$, $h(\tau)\rightarrow h(\tau)- c$ (in all other dynamic relations only $\dot{h}$ appears). Relation \eqref{repar} can be solved algebraically with respect to $h(\tau)$ and we still have intact the freedom of fixing the gauge by choosing one of the degrees of freedom as an explicit function of $\tau$.

In \cite{Igata} it was shown that conformal Killing vectors and tensors produce integrals of motion as functions of position and momenta, i.e. $I=I(x,p)$ for a massless particle. Here we see that this is true also for the massive case, with the difference that it is necessary to allow for an explicit dependence on time $I=I(\tau,x,p)$ leading to the nonlocal expression seen in \eqref{nonloc}.

\section{Integrability in terms of the nonlocal conserved charge}

In this section we want to study how the presence of a nonlocal charge may affect the notion of integrability. We saw in the previous section that such quantities possess an explicit dependence on time in terms of an integral of phase space functions. In regular systems, when we encounter time dependence in the Hamiltonian and in integrals of motion, we can consider time as a degree of freedom and thus extend the phase space dimension by two. This enhancement in the degrees of freedom does not affect integrability, because it is counterbalanced by the fact that the Hamiltonian - which previously wasn't conserved - now becomes an integral of motion.

Let us apply the procedure described above to the singular system under consideration. We assume the Hamiltonian
\begin{equation}
  H = \frac{N}{2} \mathcal{H} + u_N p_N,
\end{equation}
where $\mathcal{H}=K+1= g^{\mu\nu}p_\mu p_\nu+1\approx 0$ and $p_N\approx 0$ are the first class constraints of the theory. Additionally, we consider a linear in the momenta integral of motion of the form
\begin{equation} \label{intm1}
  I_1 = \xi_1^\alpha p_\alpha + \int\!\! \frac{N}{\Omega_1} d\tau = Q_1 +\int\!\!  \frac{N}{\Omega_1} d\tau
\end{equation}
given that $\{Q_1,H\}= \frac{N}{\Omega_1(q(\tau))} K$ holds. We introduce the gauge fixing condition
\begin{equation} \label{gfix}
  \chi = N - \Omega_1(\tau) \approx 0
\end{equation}
as an additional constraint for the theory and we write a new Hamiltonian
\begin{equation} \label{extHam}
  \bar{H} = H + u_\chi \chi.
\end{equation}
We have to add here, that we could also choose to consider the gauge fixing condition \eqref{gfix} as $\chi = N - \Omega_1(q) \approx 0$, i.e. treat the $\Omega_1$ as a function of $q$. This leads to the same results with the only difference being the in between  determined values of $u_\chi$ and $u_N$ in the process. The procedure that we follow here is the simpler one. The constraints $p_N$ and $\chi$ are now second class, since $\{p_N,\chi\} =-1$, while $\mathcal{H}\approx 0$ remains first class, i.e. $\{\mathcal{H},p_N\}= \{\mathcal{H},\chi\}=0$. The functions $u_N$ and $u_\chi$ can be evaluated through the consistency conditions (see equation \eqref{consistency}):
\begin{align}
  \dot{\chi} &\approx 0 \Rightarrow -\dot{\Omega}_1 + \{\chi, \bar{H}\} \approx 0 \Rightarrow u_N \approx \dot{\Omega}_1 \\
  \dot{p}_N & \approx 0 \Rightarrow \{p_N, \bar{H}\} \approx 0 \Rightarrow -\frac{1}{2}\mathcal{H} - u_\chi \approx0 \Rightarrow u_\chi \approx 0 .
\end{align}
The matrix $C_{\Gamma\Delta}$, $\Gamma,\Delta=1,2$ defined in \eqref{deltamat} with the help of the Poisson bracket of the second class constraints is (remember that in our case $\zeta_\Gamma = (p_N,\chi)$)
\begin{equation} \label{Cmat}
  \mathcal{C} =\begin{pmatrix}
    0 & -1 \\
    1 & 0
  \end{pmatrix} .
\end{equation}
With these results the Hamiltonian \eqref{extHam} now reads
\begin{equation}
  \bar{H} = \frac{1}{2}N \mathcal{H} + \dot{\Omega}_1\, p_N,
\end{equation}
while for the integral of motion \eqref{intm1} we obtain (by absorbing the constant of integration in $\tau$)
\begin{equation}
  I_1 = Q_1 + \tau .
\end{equation}

We now add to the phase space two more dimensions by considering $\tau$ corresponding to a dynamical degree of freedom and consider the Hamiltonian
\begin{equation}
  \tilde{H}= p_\tau + \bar{H}  = p_\tau + \frac{N}{2} \mathcal{H} + \dot{\Omega}_1(\tau)\, p_N .
\end{equation}
We notice that the following relations hold
\begin{subequations} \label{pbraint}
  \begin{align}
    \{I_1, \tilde{H}\} & \approx \mathcal{H} \approx 0 \\
    \{\mathcal{H} , \tilde{H}\} & = 0 \\
    \{I_1, \mathcal{H}\} & = \frac{2}{\Omega_1} K \neq 0.
  \end{align}
\end{subequations}
With the aid of the second class constraints  $p_N \approx 0$, $\chi \approx 0$ we define the Dirac brackets as
\begin{equation} \label{Dbraex}
  \{F,G\}_D = \{F,G\} - \{F,p_N\}\{\chi, G\} + \{F,\chi\}\{p_N, G\} .
\end{equation}
This definition is straightforward from \eqref{Dbradef} with the use of the inverse of the matrix $\mathcal{C}_{\Gamma\Delta}$ as given in \eqref{Cmat}.
With the help of \eqref{Dbraex} we can see that, concerning the Poisson brackets \eqref{pbraint}, nothing changes:
\begin{align}
  \{I_1, \tilde{H}\}_D  & = \{I_1, \tilde{H}\} \approx 0 \\
  \{\mathcal{H} , \tilde{H}\}_D & =  \{\mathcal{H} , \tilde{H}\} = 0 \\
  \{I_1, \mathcal{H}\}_D &= \{I_1, \mathcal{H}\} \neq 0.
\end{align}
For the above relations use has been made of the fact that $\{I_1,p_N\}=\{I_1,\chi\}=0$, together with the property of $\mathcal{H}$ being first class.

We can now see the difference in comparison to the regular case of non-autonomous systems where the Hamiltonian is not originally conserved. If we had considered a regular system of $d$ degrees of freedom, we would need , in order to claim Liouville integrability, $d$ independent and commuting integrals of motion. By extending the phase space when we make time a degree of freedom, we would trivially obtain the Hamiltonian as an extra (commuting with all the others) integral of motion to cover for the extra dimension added to the problem. This is not the case for a singular system describing a geodesic problem. In the latter it is possible to have integrals of motion possessing an explicit time dependence with the Hamiltonian already being autonomous. This means that when you extend the phase space adding time as degree of freedom you do not gain an extra integral of motion in terms of the Hamiltonian. The latter was already conserved before this extension (as being weakly zero). What is more, the Hamiltonian constraint $\mathcal{H}$ does not commute with $I_1$. Hence, if in such a system you have $d-1$ independent autonomous commuting integrals of motion (considering the Hamiltonian constraint as one of them), then the existence of an integral of motion like $I_1$ is not sufficient to characterize the system as Liouville integrable. However, more than one integrals of this type, like $I_1$ may exist leading to considering the system as - in principle - integrable. To make all of the above clearer, let us express it in terms of a two dimensional geodesic problem.

As we said, in a regular two dimensional system that is Liouville integrable, we have two independent integrals of motion that commute with each other. The extension of considering $\tau$ as a degree of freedom (in the case where there is an explicit time dependence) would result in a holonomic Hamiltonian that automatically provides us with an additional third integral of motion for the $2+1$ system. In the constrained case if we start from two integrals of motion one of which has an explicit time dependence and the other being already the Hamiltonian constraint (which is autonomous), the advancement to $2+1$ dimensions does not offer us a trivial integral of motion like in the regular case. We have three integrals of motion $I_1$, $\mathcal{H}$ and $\tilde{H}$ two of which, $I_1$ and $\mathcal{H}$, by definition do not commute. However, in two dimensions, we are not restricted to have only one $I_1$. We may as well choose (out of infinite possibilities) a second integral of motion linear in the momenta
\begin{equation} \label{intm2}
  I_2 = \xi_2^\alpha p_\alpha + \int\!\! \frac{N}{\Omega_2} d\tau = Q_2 +\int\!\!  \frac{N}{\Omega_2} d\tau
\end{equation}
with $\{Q_2,H\}=\frac{N}{\Omega_2(q)} K$ and satisfying the property $\{Q_1,Q_2\}=0$. After the imposition of the gauge fixing condition \eqref{gfix} the above expression reads
\begin{equation}
  I_2 =Q_2 +\int\!\!  \frac{\Omega_1}{\Omega_2} d\tau = Q_2 + f(\tau)
\end{equation}
and we obtain
\begin{equation}
  \{I_2,\tilde{H}\}_D = \{I_2,\tilde{H}\}= \dot{f} + \frac{N}{\Omega_2}K \approx \frac{\Omega_1}{\Omega_2} \mathcal{H} \approx 0 .
\end{equation}
As a result we can find two $I_1$, $I_2$ that together with the Hamiltonian $\tilde{H}$ form a Poisson algebra that satisfies
\begin{equation}
  \{I_1,I_2\} =0 , \quad \{I_i,\tilde{H}\} \approx 0, \quad i=1,2 .
\end{equation}
The existence of $I_2$ with this property is trivial in two dimensions but the practical problem that appears is that in order to know its explicit dependence on $\tau$, i.e. the functional form of $f(\tau)$, one should already be aware of the solution of the system. Hence, one could in this sense characterize every two-dimensional surface as locally integrable (given of course any necessary smoothness conditions over the metric), since at least two such commuting integrals are guaranteed to exist. Notice that this result is not in violation of known theorems setting topological obstructions to integrability of two-surfaces \cite{Kozlov,Kolokoltsov}. Since those theorems take into account quantities that are strictly functions of positions and momenta, without any possible explicit dependence on time.

Returning to our current considerations however, we may observe that the second integral $I_2$ is practically of little use in search of a solution for the system, due to the fact that you cannot turn both of them into local expressions at the same time with a single gauge fixing condition. Nevertheless, the reduction of the two dimensional geodesic problem to a first order differential equation is always possible; it is presented in the following section. For a categorization of Noetherian symmetries for generic motion of a point particle in two dimensional surfaces we refer the reader to \cite{Andr5}.

\section{The reduction of the two dimensional system}

We consider the general two dimensional metric
\begin{equation}\label{g2d}
  g_{\mu\nu} = f(x,y) \begin{pmatrix}
                        0 & 1 \\
                        1 & 0
                      \end{pmatrix}
\end{equation}
with an arbitrary function $f(x,y)$ and the local coordinates on the manifold being denoted with $x$ and $y$. Even though we assume $g_{\mu\nu}$ to be in the form of \eqref{g2d}, we do not restrict our analysis to pseudo-Riemannian spaces. A Riemannian metric can be written as in \eqref{g2d} with the help of a complex transformation. Thus, in what follows the variables $x$, $y$ could very well be complex. Only in the particular examples that we study in the next section we shall assume in general real values for the variables.

Of course, we try to treat the problem in its full generality on an arbitrary two dimensional surface; not necessarily on one with which we can associate an integral of motion commuting strongly with the Hamiltonian. That is $f(x,y)$ is not necessarily a function such that $g_{\mu\nu}$ possesses Killing vector fields or tensors. However, for any function $f(x,y)$ we know that $g_{\mu\nu}$ has an infinite number of conformal Killing vectors. Let us choose a rather simple one, namely $\xi = \frac{\partial}{\partial x}$, which has the conformal factor $\frac{\partial_x f}{f}$, i.e.
\begin{equation}
  \mathcal{L}_\xi g_{\mu\nu} = \frac{\partial_x f(x,y)}{f(x,y)} g_{\mu\nu} .
\end{equation}
We write the Lagrangian associated to the geodesic problem on \eqref{g2d} as (in comparison to \eqref{Lag} we consider here $x^\mu =(x,y)$)
\begin{equation} \label{Lag2dxy}
  L_3= \frac{1}{N} f(x,y) \dot{x} \dot{y} - \frac{N}{2}
\end{equation}
with the corresponding equations of motion
\begin{subequations}\label{eul2d}
  \begin{align}
    & f(x,y)\frac{\dot{x} \dot{y}}{N^2}+\frac{1}{2} = 0 \label{con2d}\\
    & f(x,y) \ddot{y} +\partial_y f(x,y) \dot{y}^2 - f(x,y) \frac{\dot{N}}{N} \dot{y} = 0 \label{spa12d} \\
    & f(x,y) \ddot{x} +\partial_x f(x,y) \dot{x}^2 - f(x,y) \frac{\dot{N}}{N} \dot{x} = 0, \label{spa22d}
  \end{align}
\end{subequations}
the first of which is the constraint equation of the system.

It can be easily verified that the quantity
\begin{equation}
  I = \xi^\mu \frac{\partial L_3}{\partial \dot{x}^\mu} + \int\!\! \frac{N}{2}\frac{\partial_x f(x,y)}{f(x,y)} d\tau = \frac{f(x,y)}{N} \dot{y} + \int\!\! \frac{N}{2}\frac{\partial_x f(x,y)}{f(x,y)} d\tau
\end{equation}
is conserved modulo equations \eqref{eul2d} (it is necessary to consider also \eqref{con2d} because - as we said in the previous section - these are quantities that are conserved on the constraint surface). We already have a first order relation at our hands, equation \eqref{con2d}, which is nothing but the weakly zero Hamiltonian expressed in velocity phase space variables. Let us now construct an additional first order differential equation out of $I=0$.

Let us parametrize the function $N$ in the following manner
\begin{equation} \label{reparN}
  N = \frac{2f(x,y)}{\partial_x f(x,y)} \frac{d}{d\tau} h (x,y)
\end{equation}
where $h$ is some unknown function of $x(\tau)$ and $y(\tau)$. Of course here we have to exclude from the following analysis the possibility $\partial_x f(x,y)=0$, i.e. the flat space $f(x,y)=$constant and $f(x,y)=f(y)$. In the second case the same analysis can be repeated by taking the conformal Killing vector $\eta=\partial_y$ instead of $\xi=\partial_x$. The equation $I=0$ then becomes
\begin{equation} \label{Ieq}
  \frac{\dot{y} \partial_x f}{2\left(\dot{x} \partial_x h+ \dot{y} \partial_y h \right) }+ h =0 .
\end{equation}
This is our extra first order differential equation in which we still have the freedom of choosing a particular form for $h(x,y)$ to reparametrize $N$ and fix the gauge by choosing one of the degrees of freedom to be some function of time. By solving \eqref{Ieq} with respect to $\dot{y}$, substituting into the constraint equation \eqref{con2d}, and using  \eqref{reparN} we obtain a quasi-linear first order partial differential equation for $h(x,y)$:
\begin{equation} \label{consub}
 f \partial_x h -2 h^2 \partial_y h  -h \partial_x f = 0 .
\end{equation}
Clearly if we choose the parametrization function in such a manner so that \eqref{consub} holds identically we will have the constraint equation automatically satisfied. From the theory of partial differential equations we know that the parametric solution of \eqref{consub} is given by the characteristics
\begin{equation}
  \frac{d x(s)}{ ds} = f(x(s),y(s)), \quad \frac{d y(s)}{ ds} = -2 h(s)^2, \quad \frac{d h(s)}{ ds} = h(s) \partial_x f(x(s),y(s)) .
\end{equation}
Given smooth enough conditions on $f(x,y)$ the above system always has a solution (at least locally). It may happen however that this solution is not unique. Each choice satisfying \eqref{consub} sets a particular gauge condition for which the solution of the full system holds.

Since both functions $f$ and $h$ are arbitrary at this point, it is easier to go the other way round and parametrize function $f$ with respect to $h$ so that \eqref{consub} is satisfied. This leads to
\begin{equation} \label{parf}
  f(x,y) = -2 h(x,y) \left( \int\!\! \partial_y h(x,y) dx + h_1(y)\right)
\end{equation}
with $h(x,y)$ remaining arbitrary and $h_1(y)$ the integration function due to the presence of an indefinite integral.

With the help of \eqref{reparN}, \eqref{Ieq} and \eqref{parf} the original system \eqref{eul2d} is expressed as
\begin{subequations} \label{almostsol}
\begin{align} \label{almostsol1}
  N & = 2 h(x,y) \dot{x} \\ \label{almostsol2}
  \frac{\dot{y}}{\dot{x}} & = \frac{h(x,y)}{\int\!\! \partial_y h(x,y) dx + h_1(y)} .
\end{align}
\end{subequations}
It seems that there is some arbitrariness residing in the form of $h_1(y)$ in the denominator of \eqref{almostsol2}, and of course in the right hand side of \eqref{parf}. But this is not so. In reality this ``arbitrariness" can be lifted and $h_1(y)$ can be determined algebraically by demanding that the given $f(x,y)$ satisfies \eqref{parf} for any $h(x,y)$ solving \eqref{consub}. We shall see exactly how this works in the examples that follow in the next section. It can be easily verified that relations \eqref{almostsol} satisfy \eqref{eul2d}. We want to notice that we are essentially left with only a single first order equation, namely \eqref{almostsol2}; since we can consider \eqref{almostsol1} as merely prescribing $N$. Either of $x$ or $y$ may serve in \eqref{almostsol2} as the time variable, thus fixing the gauge. For every function $h(x,y)$, corresponding to a conformal factor $f(x,y)$, relations \eqref{almostsol} provide the solution of the system. In particular problems however, the function $f(x,y)$ is known, so in order to make use of \eqref{almostsol} one needs to determine the corresponding function $h(x,y)$. In the following section we study a few examples of such cases, but first we shall demonstrate by means of a counter example that this reduction is not the effect of some point symmetry.

\subsection{The reduction as not the effect of a point symmetry}

The reduction of the system, through the nonlocal conserved charge, to the first order equation \eqref{almostsol2} is not trivial and  it can take place for every smooth enough function $f(x,y)$. Even though the corresponding $h(x,y)$ might be difficult to be found. We want to show that this property is not connected  to a possible existence of some Lie-point symmetry. It can be easily verified that there exist functions $f(x,y)$ for which no Lie-point symmetry is present for the system. However, the reduction to \eqref{almostsol2} is in principle always possible due to the existence of the nonlocal conserved charge and of the solution \eqref{parf} of \eqref{consub}.

In order to demonstrate this fact we can follow the subsequent procedure: Solve the constraint equation \eqref{con2d} with respect to the auxiliary degree of freedom $N$ and substitute to the  two equations \eqref{spa12d} and \eqref{spa22d}. Then, we see that the latter become a single second order equation involving two degrees of freedom $x(\tau)$ and $y(\tau)$. The gauge freedom of the system allows us to consider one of the two as time. Let us choose $x(\tau)=\tau$, then this equation reads
\begin{equation} \label{notLeq}
  \ddot{y}=\frac{\dot{y}}{f(\tau,y)}\left(\partial_\tau f(\tau,y) - \dot{y} \partial_y f(\tau,y)\right).
\end{equation}
The existence of a Lie-point symmetry of \eqref{notLeq} that would lead to the reduction to a first order equation now depends on the function $f(x,y)$ (or $f(\tau,y)$ in its gauge fixed version $x=\tau$). We are not going to expatiate upon the theory of point symmetries, for which we refer the reader to the well known textbooks \cite{Steph,Olver}. We just confine ourselves to say that in order for a point symmetry generator $X=\eta(\tau,y)\partial_\tau + \phi(\tau,y)\partial_y$ to exist for equation \eqref{notLeq}, the following system of linear partial differential equations must have some solution
\begin{subequations} \label{Liesys}
\begin{align}
  & \partial_{y,y}\eta-\frac{\partial_y f \partial_y \eta}{f} =0 \\
  & \partial_{\tau,\tau}\phi -\frac{\partial_\tau f \partial_t \phi}{f} =0\\
  & 2 \partial_\tau f  \partial_y \eta + \frac{ \partial_y f \left(\eta \partial_\tau f + \phi \partial_y f  \right)}{f}- \eta \partial_{\tau,y}f  - \partial_y f \partial_y \phi - \phi\partial_{y,y}f -f \left(\partial_{y,y}\phi -2 \partial_{\tau,y}\eta \right) = 0\\
  & \partial_\tau f \partial_\tau \eta - \frac{\partial_\tau f \left(\eta \partial_\tau f + \phi \partial_y f \right)}{f}+ \eta \partial_{\tau,\tau}f -2 \partial_y f \partial_\tau \phi + \phi \partial_{\tau,y}f - f \left(2 \partial_{\tau,y}\phi - \partial_{\tau,\tau}\eta \right) = 0
\end{align}
\end{subequations}
with respect to $\eta(\tau,y)$ and $\phi(\tau,y)$. The function $f(\tau,y)$ is considered known. It can be seen that \eqref{Liesys} does not have a solution for every possible $f(\tau,y)$ that one may consider. For example, if we assume a two dimensional surface having $f(x,y)=e^{x y (x+y)}$, the system \eqref{Liesys} leads to an incompatibility. Hence, no Lie-point symmetry exists. On the other hand, the reduction to \eqref{almostsol2} is achieved for an arbitrary function $h(x,y)$ and thus $f(x,y)$. As a result, we see that the aforementioned reduction is not related to a point symmetry and the nonlocal conserved charges offer a new non-conventional way of simplifying a dynamical system. In the following section we are going to study several examples making use of this property for the geodesic equations on two dimensional surfaces.

\section{Simple examples of two dimensional systems}

In this section we examine a few examples, starting from a trivially simple and moving on to more complicated ones.

\subsection{Example 1. The flat space}

In order to demonstrate how the above considerations work, let us consider for example the trivial case
\begin{equation}
  f(x,y) = x,
\end{equation}
which results in a two dimensional flat space. The, equation \eqref{consub} implies that $h(x,y)$ is any function for which
\begin{equation} \label{arbF}
  F\left(\frac{h(x,y)}{x}, h(x,y)^2+y\right) =0,
\end{equation}
where $F$ is some function of its arguments. The important thing in order to obtain the general solution for the geodesics is to involve both branches. Let us take the simpler possibility, the linear combination
\begin{equation} \label{exampF1}
  F = c_1 \frac{h(x,y)}{x}+ c_2 \left( h(x,y)^2+y \right)=0
\end{equation}
where $c_1, c_2$ are constants. Obviously, the solution of the above equation depends on one constant because we can reparametrize $c_1 = 2 \kappa_1 c_2$ and divide by $c_2$. Notice that the constant $c_2$, unlike $c_1$, cannot be zero because it would lead to $h(x,y)=0$ which is excluded from our analysis. Thus, we obtain from \eqref{exampF1}
\begin{equation} \label{ex1h}
  h(x,y) = \frac{-\kappa_1 \pm \sqrt{\kappa_1^2-x^2 y}}{x} .
\end{equation}
At this point the system \eqref{almostsol} implies
\begin{align}
  N & = \frac{2 \dot{x} \left(\sqrt{\kappa_1^2- x^2 y}-\kappa_1\right)}{x} \\ \label{almsolex}
  \frac{\dot{y}}{\dot{x}} & = \frac{2 y\left(\sqrt{\kappa_1^2- x^2 y}-\kappa_1\right)}{x \left(\sqrt{\kappa_1^2- x^2 y}-2 y h_1(y)\right)}
\end{align}
where $h_1 (y)$ is some function of integration whose explicit dependence can be found algebraically by demanding that \eqref{parf} holds for $f(x,y)=x$. Substitution of the latter and of \eqref{ex1h} in \eqref{parf} leads to
\begin{align*}
  & x = \frac{\left(\pm \sqrt{\kappa_1^2-x^2 y}-\kappa_1\right) \left(2 y h_1(y) \mp \sqrt{\kappa_1^2-x^2 y}\right)}{x y} \Rightarrow \\
  & \frac{\left(\sqrt{\kappa_1^2-x^2 y} \mp \kappa_1\right) }{x y} \left(\kappa_1+2 y h_1(y)\right)= 0,
\end{align*}
which implies $h_1(y) = -\frac{\kappa_1}{2 y}$, thus fixing the function $h_1(y)$ appearing in \eqref{parf}. Now equation \eqref{almsolex} can be straightforwardly integrated. In the gauge $x(\tau)=\tau$ we may express the final result as
\begin{align}
  N(\tau) & = -2 \kappa_2 \tau \\
  y(\tau) & = \kappa_2 \left(2\kappa_1-\kappa_2 \tau^2\right)
\end{align}
which is the general solution of the geodesic equations in this gauge. It can easily be seen that a different choice of a function $F$ in place of \eqref{exampF1} results essentially in the same solution with a different parametrization in what regards the constants of integration.

\subsection{Example 2. Space of constant Ricci scalar}

If we choose
\begin{equation} \label{ex2af}
  f(x,y) = -\frac{4}{R (x+y)^2},
\end{equation}
where $R$=const. we have a space of constant scalar curvature, $R$. Of course we can normalize the constant to be plus or minus one, but in order to deal at the same time with both cases we just leave is as it is. For this choice of $f(x,y)$, equation \eqref{consub} assumes the form
\begin{equation}
  h^2 \partial_y h -\frac{ 2}{R (x+y)^2} \partial_x h -\frac{4}{R (x+y)^3} h = 0 .
\end{equation}
The general solution to this equation is given by those functions $h(x,y)$ for which
\begin{equation}
  F\left( \frac{R (x+y)^2 h(x,y)^2 - 2}{(x+y)^2 h(x,y)}, \frac{R x (x+y)^2 h (x,y)^2+2 y}{R (x+y)^2 h (x,y)^2 - 2} \right) =0.
\end{equation}
We choose the following combination including both branches of the solution
\begin{equation} \label{consubex2a}
 F= 2 \kappa_1 \left(\frac{R (x+y)^2 h(x,y)^2 - 2}{(x+y)^2 h(x,y)}\right)^{-1} + \left( \frac{R x (x+y)^2 h (x,y)^2+2 y}{R (x+y)^2 h (x,y)^2 - 2} \right) =0,
\end{equation}
where $\kappa_1$ is a constant introduced in the same manner as in the previous example. Equation \eqref{consubex2a} results in
\begin{equation}\label{ex2ah}
  h(x,y) = -\frac{\kappa_1}{R\,  x} \pm \frac{\left[(x+y)^2\left(\kappa_1^2 \left(x+ y\right)^2-2  R x y\right)\right]^{1/2}}{R\,  x  (x +y)^2}.
\end{equation}

Use of the latter together with \eqref{ex2af} inside \eqref{parf} leads to the determination of the integration function $h_1(y)$ which is
\begin{equation}
  h_1(y) = - \frac{\kappa_1}{R\, y}.
\end{equation}
The set \eqref{almostsol} becomes
\begin{align} \label{exprN}
  N & = -2 \dot{x} \left[ \frac{\kappa_1 }{R\,  x} \mp \frac{\left[(x+y)^2\left( \kappa_1^2 \left(x+y\right)^2 - 2 R x y \right)\right]^{1/2}}{R\,  x  (x +y)^2}\right] \\ \label{ex2ay}
  \frac{\dot{y}}{\dot{x}} & = \frac{\kappa_1^2 (x +y)^2- R  x y \mp \kappa_1  \sqrt{(x+y)^2\left(\kappa_1^2 (x+y)^2-2 R  x y\right)}}{R x^2} .
\end{align}
The last equation can be easily integrated to give
\begin{equation} \label{solsph}
  y_{\pm} = \frac{\kappa_2^2 R  x\mp \sqrt{2} \sqrt{\kappa_1^2 \kappa_2^2 \left(2 \kappa_1^2- R \right) \left(\kappa_2^2-x^2\right)^2}}{2 \kappa_1^2 \kappa_2^2+\left(R -2 \kappa_1^2\right) x^2}.
\end{equation}
We have to note here, that the $y_{\pm}$ solution results for each of the two plus and minus equations \eqref{ex2ay} corresponding to each of \eqref{ex2ah}, creating a totality of four combinations. Depending on the range of the parameters $\kappa_1$, $\kappa_2$, $R$ and the variable $x$ the combination that represents the solution of the system changes. The solution is extracted in an arbitrary gauge, we can easily set $x(\tau)$ as some explicit function of the time parameter $\tau$ in \eqref{exprN} and \eqref{ex2ay} and have the trajectories expressed with respect to that specific parameter.

To compare with what we know from the theory of spherical surfaces, we may use the transformation
\begin{subequations} \label{trantosph}
\begin{align}
  x & = \cot \left[\frac{\sqrt{R} }{2 \sqrt{2}} \phi  + \frac{1}{2} \ima \ln \left[\cot \left(\frac{\sqrt{R }}{2 \sqrt{2}} \theta  \right)\right]\right]\\
  y & = \tan \left[\frac{\sqrt{R}}{2 \sqrt{2}}  \phi  -\frac{1}{2} \ima \ln \left[\cot \left(\frac{\sqrt{R}}{2 \sqrt{2}} \theta   \right)\right]\right]
\end{align}
\end{subequations}
that makes the two dimensional metric become
\begin{equation} \label{2dsph}
  g_{\mu\nu}= \begin{pmatrix}
                1 & 0 \\
                0 & \sin^2 \Theta
              \end{pmatrix},
\end{equation}
where $\Theta=\frac{\sqrt{R}}{\sqrt{2}}\theta$. In the gauge $N=$constant, and for $\Theta = \frac{\pi}{2}$, we expect a linear solution in $\phi$. Truly, if we consider the Lagrangian \eqref{Lag} with the metric \eqref{2dsph}, its Euler-Lagrange equations become
\begin{align}
  & \frac{\dot{N}}{N} \dot{\Theta} +\frac{R}{2}\sin (\Theta) \cos (\Theta) \dot{\phi}^2 - \ddot{\Theta}=0 \\
  & -\frac{\dot{N}}{N} \dot{\phi} + 2 \cot (\Theta) \dot{\Theta} \dot{\phi} + \ddot{\phi} =0 \\
  & \frac{2}{R\, N} \dot{\Theta}^2 +\frac{\sin ^2(\Theta) }{N} \dot{\phi}^2+ N = 0
\end{align}
and it is easy to see that the configuration
\begin{equation}
  N=1, \quad \Theta = \frac{\pi}{2}, \quad \phi=\ima\, \tau
\end{equation}
is a solution. The imaginary unit in $\phi$ has to do with the fact that we took the non-kinetic term in \eqref{Lag} as $-N/2$ with the metric being positive definite. Had we considered $+N/2$ instead, the solution would be $\phi=\tau$. However, since we just want to see the analogy with the specific example under consideration, we choose to keep the same $-N/2$ part in the Lagrangian. From the transformation \eqref{trantosph} we can verify that this solution corresponds to
\begin{equation} \label{solsph2}
  x= - \ima \coth \left(\frac{\sqrt{R} }{2 \sqrt{2}}\tau\right), \quad y = \ima \tanh \left(\frac{\sqrt{R} }{2 \sqrt{2}}\tau \right).
\end{equation}
The imaginary units again have to do with the fact that we map a solution for a positive definite metric \eqref{2dsph} to one of a Lorentzian signature \eqref{g2d} (with $f(x,y)$ given by \eqref{ex2af}). From \eqref{solsph} we notice that for the particular values $\kappa_1=0$, $\kappa_2=1$ of the integration constants, we obtain
\begin{equation}
  y= \frac{1}{x}
\end{equation}
which we immediately see that is compatible with \eqref{solsph2}. What is more, if we set \eqref{solsph2} back into \eqref{exprN} - again for $\kappa_1=0$ and $\kappa_2=1$ - we obtain $N=\pm 1$ as expected. Thus, we observe that this particular result is contained in the full solution that we managed to extract in an arbitrary gauge.

\subsection{Example 3}

Let us proceed with a less trivial system.To that end we consider the more complicated conformal factor
\begin{equation} \label{ex3f}
  f(x,y) = - x^3 e^{y} \left(x+e^{y}\right) .
\end{equation}
It can be checked that the corresponding metric \eqref{g2d} does not possess a Killing vector or tensor up to second order apart from itself. In this case it is not straightforward to integrate equation \eqref{consub}. However, we can see that the function $h(x,y)= x e^{y}+x^2$ is a partial solution. Thus, we can at least recognise an integrable sector that gives us explicitly a solution to the geodesic equations. By following the same procedure as before \eqref{almostsol} result in (the corresponding $h_1(y)$ function, that is obtained algebraically by checking the consistency of \eqref{parf}, is zero in this case)
\begin{subequations} \label{sol}
\begin{align}
  N(\tau) & =2 \tau^2 (c_1 \tau-1) \\
  x(\tau) & = \tau \\
  y(\tau) &= \ln \left[ \tau^2 \left(c_1-\frac{2}{\tau}\right) \right],
\end{align}
\end{subequations}
with $c_1$ being the constant of integration. It can be seen that solution \eqref{sol} together with $x=\tau$, satisfies the system \eqref{eul2d}. The reason for having only one constant of integration in this solution, instead of two in the previous example, is because we used a partial solution $h(x,y)$ of \eqref{consub} and not the more general that exists. However, and even though, the system does not have an autonomous integral of motion (at least up to second order), we were able to obtain a partial solution with the help of the integration based on the nonlocal conserved charge.

In order to demonstrate how difficult it can prove in many cases to find solutions given in terms of elementary functions we need only study the resulting equation \eqref{consub} with the $f(x,y)$ given by \eqref{ex3f}. As we discussed, the $h(x,y)= x e^{y}+x^2$, given above is a particular solution of \eqref{consub}. We may notice that for the resulting equation
\begin{equation} \label{consubex3}
  x^3 e^y \left(x+e^y\right) \frac{\partial_x h}{h}+2 h  \partial_y h-x^2 e^y \left(4 x+3 e^y\right) =0 ,
\end{equation}
a more general group invariant solution can given by introducing a new function $h_1(e^y/x)$ if we set $h(x,y)= x^2 \left(\frac{e^y}{x}+1\right) \sqrt{\frac{1}{2} \left(h_1\left(e^y/x\right)+1\right)}$. Then, \eqref{consubex3}, reduces to the ODE
\begin{equation}
  \left[(s+1) h_1(s)+1\right] \frac{d}{ds}h_1(s) + 2 \left(h_1(s)^2-1\right) =0,
\end{equation}
where $s=\frac{e^y}{x}$ and where we also see why the previous particular solution, that corresponds to $h_1(s)=1$, worked. Of course, in order to differentiate from before, we require $h(s)\neq \pm 1$. Under this condition, the solution of the above equation can be provided algebraically by solving
\begin{equation}
  \frac{\kappa_1}{\left(h_1^2-1\right)^{1/4}} + \, _2F_1\left(\frac{3}{4},1;\frac{3}{2};h_1^2\right) h_1 -2 (s+1)=0,
\end{equation}
where $\kappa_1$ is a constant of integration and $_2F_1$ is the Gauss hypergeometric function. Obviously, this solution cannot be practically useful to obtain the corresponding geodesic equations. A thing that proves still the difficulties that lie even within the aforementioned reduction.

\subsection{Example 4. A non-integrable pseudo-Euclidean Toda system}

There is a lot of work dedicated to Euclidean Toda and generalized Toda systems. The same is not true for the pseudo-Euclidean case where the metric of the kinetic part has a Lorentzian signature. However, there exist some known integrable cases of such systems \cite{Gavrilov}. In two dimensions a generalized pseudo-Euclidean ($\tilde{g}_{\mu\nu} =\mathrm{diag}(-1,1)$) Toda system with a two part contribution in the potential can be written as (we present the parametrization invariant version of system)
\begin{equation} \label{LagToda}
  L = \frac{1}{2 n} \tilde{g}_{\mu\nu} U^\mu U^\nu - \frac{n}{2} \left(a_1 e^{b^{(1)}_\mu U^\mu} + a_2 e^{b^{(2)}_\mu U^\mu} \right),
\end{equation}
where we assume $b^{(1)}$ and $b^{(2)}$ to be both nonzero, linearly independent vectors on $\mathbb{R}^2$. The conditions for such a system to be integrable are known \cite{Gavrilov,Mel}, $b=b^{(1)}-b^{(2)}$ needs to be an isotropic vector, i.e. $b^\mu b_\mu=0$. In coordinates $U^\mu=(u,v)$ the most general potential we can write is
\begin{equation}
  V(u,v) = a_1 e^{l_1 u+l_2 v}+ a_2 e^{l_3 u+l_4 v},
\end{equation}
with $l_1,l_2,l_3,l_4$ arbitrary constants. Here, we choose to study a simpler version which however still does not belong to a known integrable class. To that end we take
\begin{equation} \label{l1ass}
  l_1 = 2\left(l_3+l_4\right)- l_2 .
\end{equation}
It is easy to verify that, under \eqref{l1ass}, the difference of the two vectors, $b^\mu =(l_2-l_3-2l_4,l_2-l_4)$, is nonzero and orthogonal to itself only if $l_3=-l_4$ or $l_3 = 2l_2-3l_4$. These are the values that correspond to the known integrable class. We shall refrain for such an assumption and consider $l_2,l_3$ and $l_4$ as completely arbitrary. Hence, we deal with a pure non-integrable case.

In order to use the expression we proved in the previous section, we perform the reparametrization $n\mapsto N = n V(u,v)$ and at the same time we adopt the coordinate transformation
\begin{equation}
  u = \frac{x-y}{\sqrt{2}}, \quad v = \frac{x+y}{\sqrt{2}},
\end{equation}
so that the Lagrangian \eqref{LagToda} becomes \eqref{Lag2dxy} with
\begin{equation} \label{ex4f}
  f(x,y) = a_1 e^{\sqrt{2} ((l_3+l_4) x+y (l_2-l_3-l_4))} + a_2 e^{\frac{(l_3+l_4) x+(l_4-l_3) y(t)}{\sqrt{2}}} .
\end{equation}
Once more it is not trivial to integrate \eqref{consub}, to find the corresponding $h(x,y)$. However, a partial solution can be easily derived and it has the form
\begin{equation} \label{ex3h}
  h(x,y) = \left( \frac{a_1 (l_3+l_4)}{2 (l_3+l_4-l_2)}\right)^{1/2} e^{\frac{(l_3+l_4)x (l_2-l_3-l_4) y}{\sqrt{2}}} .
\end{equation}
Now, \eqref{almostsol} can be written as
\begin{subequations} \label{almostsolToda}
\begin{align}
  N & = \left( 2 \frac{a_1 (l_3+l_4)}{(l_3+l_4-l_2)}\right)^{1/2} e^{\frac{(l_3+l_4)x (l_2-l_3-l_4) y}{\sqrt{2}}} \dot{x}  \\ \label{almostsolToda2}
  \frac{\dot{y}}{\dot{x}} & = \left[\frac{l_2}{l_3+l_4}-1-\left( \frac{2 (l_3+l_4-l_2)}{a_1 (l_3+l_4)}\right)^{1/2} e^{-\frac{l_2 y(t)+(l_3+l_4) (x-y)}{\sqrt{2}}} h_1(y) \right]^{-1} .
\end{align}
\end{subequations}
Substitution of \eqref{ex3h} and \eqref{ex4f} into \eqref{parf} results into the algebraic determination of the integration function $h_1(y)$ which reads
\begin{equation} \label{h1Toda}
  h_1 (y) = a_2 \left(\frac{l_3+l_4-l_2}{2 a_1 (l_3+l_4)}\right)^{1/2} e^{-\frac{(l_2-2 l_4) y}{\sqrt{2}}}.
\end{equation}
It is more convenient now to integrate \eqref{almostsolToda2} with respect $x(\tau)$ which results into
\begin{equation}
  x(\tau) = \frac{\sqrt{2}}{l_3+l_4} \ln\left[c_1 + \frac{a_2 (l_3+l_4-l_2)}{a_1 (3 l_2-2 (l_3+2l_4))} e^{\frac{(2 l_3+4l_4-3 l_2)y(\tau) }{\sqrt{2}}}\right] + \frac{l_2-l_3-l_4}{l_3+l_4} y(\tau) .
\end{equation}
Finally, we can express the solution in the gauge $y(\tau)=\tau$ as
\begin{align}
  N(\tau) & = -\left(\frac{l_3+l_4-l_2}{a_1 l_3+l_4}\right)^{1/2}\left[\frac{\sqrt{2} a_2 (l_3+3 l_4-2 l_2) e^{-\frac{(l_2-2 l_4)\tau}{\sqrt{2}}}}{ (2l_3+4 l_4-3 l_2)} + c_1 e^{\sqrt{2} (l_2-l_3-l_4)\tau} \right]\\
  x(\tau) & = \frac{\sqrt{2}}{l_3+l_4} \ln\left[c_1 + \frac{a_2 (l_3+l_4-l_2)}{a_1 (3 l_2-2 (l_3+2l_4))} e^{\frac{(2 l_3+4l_4-3 l_2)\tau }{\sqrt{2}}}\right] + \frac{l_2-l_3-l_4}{l_3+l_4} \tau \\
  y(\tau) & = \tau .
\end{align}
As we see, the solution entails one integration constant, that is owed to the fact that we were able to only determine a partial solution of \eqref{consub}. Nevertheless, we managed - with the use of a nonlocal conserved charge - to extract a partial solution in terms of elementary functions for the equivalent geodesic problem of a system which originally does not belong in a known integrable class.

\section{Conclusion}

In this work we studied the system of geodesic equations in its full form. By keeping the parametrization invariance intact, we defined nonlocal conserved charges in phase space with the help of conformal Killing vectors of the manifold metric. These quantities are conserved modulo the constraint equation of motion. This is why the gauge invariance of the system is important for their realization. Due to the fact that these integrals of motion posses an explicit time dependence we considered time as a dynamical variable and extended accordingly the Hamiltonian function, so as to examine the effect of such conserved quantities in the integrability of the system.

Due to the fact that two dimensional surfaces have an infinite number of conformal Killing vectors, it can be shown that in principle there exist enough commuting conserved quantities in phase space, so as to characterize the system as Liouville integrable. The drawback is however, that one cannot know the explicit dependence in time for all these rheonomic integrals of motion. This practically means that the explicit transformation to the action angle variables cannot be known. Even so, we demonstrated that the corresponding system can always be reduced to a single first order differential equation. The latter being the effect of the existing nonlocal charges and not of an underlying Lie-point symmetry, since it was done for an arbitrary surface. We studied some particular examples in order to demonstrate how this method works: from flat space to a generalized pseudo-Euclidean Toda system that does not belong to a known integrable class and which we associated to a corresponding geodesic problem.

Apart from geodesic problems, these nonlocal conserved charges can be used in more general parametrization invariant Lagrangians involving a potential function. This has been done in cosmological problems in order to derive new solutions \cite{FLRWgen} as well as more recently to prove the integrability of the mixmaster model \cite{intIX}.

As happens with conventional point symmetries - corresponding to Killing vectors for a geodesic problem - which can be generalized to symmetries involving Killing tensors, the same can be done with the nonlocal conserved charges. We can write in a similar manner the conserved charge \cite{confp}
\begin{equation} \label{highordnonloc}
  I = \Xi^{\kappa \alpha_1 ...\alpha_n} p_\kappa p_{\alpha_1} ... p_{\alpha_n} + \int\!\! N \omega^{\alpha_1...\alpha_n} p_{\alpha_1} ... p_{\alpha_n} d\tau ,
\end{equation}
which is of $n+1$ order in the momenta and constructed out of conformal Killing tensors
\begin{equation}
  \nabla^{(\mu}\Xi^{\nu \alpha_1 ...\alpha_n)} = \omega^{(\alpha_1 ...\alpha_n} g^{\mu\nu)}
\end{equation}
of the manifold metric $g_{\mu\nu}$. Again we see how for Killing tensors \eqref{highordnonloc} reduces to the usual local expression $\Xi^{\kappa \alpha_1 ...\alpha_n} p_\kappa p_{\alpha_1}$. The nonlocal part, involving the integral of phase space quantities, can be considered as a pure function of the time variable on the solution of the system. There is of course the issue of the possible existence of a solution so that this integral makes sense. This is a matter of considering a smooth enough metric, so that the relevant existence theorems of ordinary differential equations can guarantee this fact for the geodesic system.

In all, we have observed that the parametrization invariance, when present, can be exploited in such a manner so as to gain an insight in the integration of the system of equations at hand. Given specific gauge choices, like for example \eqref{gfix} in \eqref{intm1}, an additional first order relation of the form $\xi^\alpha p_\alpha + \tau =0$ can be obtained to be used together with the equations of motion. Due to its rheonomic nature is independent to other first order relations provided by autonomous integrals of motion. Thus, the opportune strategy is not to fix the gauge of a system blindly, before studying the symmetry structure of the problem.

\end{document}